\documentclass[journal]{IEEEtran}

\usepackage{amsmath,amsfonts}
\usepackage{algorithmic}
\usepackage{array}
\usepackage{textcomp}
\usepackage{stfloats}
\usepackage{url}
\usepackage{verbatim}
\usepackage{graphicx}
\usepackage{amssymb}
\usepackage{color,soul}
\usepackage{amsthm} 
\usepackage{bbm}
\usepackage{algorithm}
\usepackage{caption}
\usepackage{bm}
\usepackage{subcaption}
\usepackage{xpatch}
\usepackage{geometry}
\def\BibTeX{{\rm B\kern-.05em{\sc i\kern-.025em b}\kern-.08em
    T\kern-.1667em\lower.7ex\hbox{E}\kern-.125emX}}

\begin{document}
\title{Channel Estimation for Holographic MIMO: Wavenumber-Domain Sparsity Inspired Approaches}

\author{Yuqing~Guo,~Yuanbin~Chen,~and~Ying~Wang,~\IEEEmembership{Member,~IEEE}

\thanks{
This work was supported in part by the Beijing Natural Science Foundation under Grant 4222011. \textit{(Corresponding author: Ying Wang.)}

The authors are with the State Key Laboratory of Networking and Switching Technology, Beijing University of Posts and Telecommunications, Beijing 100876, China  (e-mail:guoyuqing2020@bupt.edu.cn; chen\_yuanbin@163.com; wangying@bupt.edu.cn).

}

\vspace{-1em}
}


\maketitle

\begin{abstract}

This paper investigates the sparse channel estimation for holographic multiple-input multiple-output (HMIMO) systems.
Given that the wavenumber-domain representation is based on a series of Fourier harmonics that are in essence a series of orthogonal basis functions, a novel wavenumber-domain sparsifying basis is designed to  expose the sparsity inherent in HMIMO channels. Furthermore, by harnessing the beneficial sparsity in the wavenumber domain, the sparse estimation of HMIMO channels is structured as a compressed sensing problem, which can be efficiently solved by our proposed wavenumber-domain orthogonal matching pursuit (WD-OMP) algorithm. Finally, numerical results demonstrate that the proposed wavenumber-domain sparsifying basis maintains its detection accuracy regardless of the number of antenna elements and antenna spacing. Additionally, in the case of antenna spacing being much less than half a wavelength, the wavenumber-domain approach remains highly accurate in identifying the significant angular power of HMIMO channels.

\end{abstract}


\section{Introduction}

Holographic multiple-input multiple-output (HMIMO) has been envisaged as a viable candidate for future MIMO technologies. Characterized by its nearly continuous antenna aperture, HMIMO is anticipated to introduce an unprecedented degree of freedom (DoF) in the manipulation of electromagnetic (EM) waves, facilitating the super-directivity beam having a low sidelobe leakage \cite{Holo-26} with reduced power consumption \cite{zongshu}, which can also be applied to computational imaging \cite{update3}. To fully harness the beneficial beamforming capabilities of HMIMO, the accurate acquisition of channel state information (CSI) is imperative.

At present, a few fledgling efforts have been devoted to HMIMO channel estimation based on different channel models that are informally bifurcated into two primary categories. { The first kind includes EM channel model, which is deduced from Maxwell equations essentially \cite{update1}}, \cite{fourier,spatialcharacter}. The second category is parametric physical channel based on multi-path representation \cite{chen-jsac3,holographicchannelestimation1}. To be explicit, the former goes more deeply into capturing the essence of EM propagation in arbitrary scattering environment. The latter is more concentrated on the imposed sparsity determined by the significant paths associated with clusters in the scattering environment. Based on the EM channel model, in \cite{holographicchannelestimation2}, HMIMO channel is estimated using a least-squares (LS) estimator in the absence of prior information of the channel statistics. By contrast, as for parametric physical channels, angular and distance parameters associated with HMIMO channels are accurately retrieved in \cite{chen-jsac3,holographicchannelestimation1} through the use of various tricks. Despite utilizing different analytical representations, the EM channel model and the multi-path channel model converge on the same underlying principle, i.e. the multi-path channel model can be transformed into its EM counterpart based on the Weyl's Indentity \cite{spatialcharacter}.


For the high-dimensional channel matrix resulting from a large number of antenna elements, we always strive to figure out their potential sparsities to reduce the overhead and complexity when carrying out channel estimation. For instance, by harnessing the angular-domain sparsity in massive MIMO (mMIMO) channels \cite{AD}, the overhead is exclusively dependent on the number of clusters within the propagation environment rather than the number of antenna elements. This inspires us to explore the potential sparsity inherent in HMIMO channels. Given that EM wave propagation can always be decomposed into a series of plane
waves propagating along different directions, the DoFs of which depend on the array sizes rather than the number of antenna elements, we shift our focus to the wavenumber-domain representation as specified in \cite{fourier}.
While the terminology  ``wavenumber" has been present for quite a while, there have been limited efforts on examining the sparsity of HMIMO channels in the wavenumber domain, particularly for their efficient channel estimates.
Therefore, the goal of this paper set out with the goal of filling this knowledge gap in the state-of-the-art by following contributions. 

Firstly, we consider an HMIMO system consisting of a transmitter and a receiver, both equipped with a uniform planar array (UPA). The channel between the transceivers is characterized by the EM-compliant Fourier plane-wave series expansion for encapsulating the arbitrary scattering environment. Given that the wavenumber-domain representation is based on a series of Fourier harmonics (FHs), which are in essence a series of orthogonal basis functions, a novel wavenumber-domain sparsifying basis is crafted to restructure the original HMIMO channel as a sparse one. 
Furthermore, the proposed wavenumber-domain sparsifying basis maintains its detection accuracy independent of the number of antenna elements and antenna spacing, in sharp contrast to the angular-domain sparsifying basis.
Secondly, by harnessing the sparsity in the wavenumber domain, the channel estimation is structured as a compressed sensing (CS) problem. To solve this problem, a wavenumber-domain orthogonal matching pursuit (WD-OMP) algorithm is proposed to efficiently estimate sparse HMIMO channels. Finally, simulation results substantiate the effectiveness of the proposed algorithm. We demonstrate that 
in the case of antenna spacing being much less than half a wavelength, the wavenumber-domain technique remains highly efficient in accurately identifying the HMIMO channels, revealing its great robustness.

\section{System Model}
We consider an HMIMO system operating at central frequency of $f_c$, consisting of both a transmitter and a receiver, each equipped with a UPA. The array sizes of the UPAs are represented by $L_{S,x} \times L_{S,y}$ for the transmitter and $L_{R,x} \times L_{R,y}$ for the receiver, respectively. The antenna spacing for both UPAs is uniformly maintained at $\delta$. The number of antenna elements is denoted by $N_S = N_{S,x}\times N_{S,y}$ for the transmitter and $N_R = N_{R,x}\times N_{R,y}$ for the receiver, where both $N_{I,x}$ and $N_{I,y} $ ($I\in \left\{S,R\right\}$) are odd, indicating the number of antenna elements along the $x$-axis and $y$-axis, respectively. The index of each antenna element is denoted by $n_I\triangleq\left(n_{I,x}, n_{I,y}\right), I \in \left\{S,R\right\}$, where $n_{I,i} \in \left\{\frac{1-N_{I,i}}{2},\dots,0,\dots, \frac{N_{I,i}-1}{2}\right\}, i\in \left\{x,y\right\}$ indicates the antenna index along the $x$- and $y$-axis, respectively. Furthermore, both the transmitter and the receiver coincide with the horizontal $xOy$ surface with Cartesian coordinates ${\bf{p}}_S = \left[0, 0, 0\right]$ and ${\bf{p}}_R = \left[x_0,y_0, z_0\right]$, without loss of generality. Thus, the location of the $n_{I}$-th antenna can be given by ${\bf p}_{I,n_I}={\bf p}_{I} +\delta\left[n_{I,x},n_{I,y},0\right]^T,I\in \left\{S,R\right\}$, where $n_I = 1 + (n_{I,x} + \frac{N_{I,x}-1}{2}) \times N_{I,y} + n_{I,y} + \frac{N_{I,y}-1}{2}$.

\subsection{{Fourier-Harmonic-Based Channel Representation }}\label{subsec2}
In accordance with the Fourier plane-wave series expansion~\cite{fourier}, the point-to-point channel response $h\left({\bf s}, {\bf r}\right)\in \mathbb{C}$, emanating from the transmitter location $\mathbf{s}=\left[s_x,s_y,s_z\right]$ to the receiver location $\mathbf{r}=\left[r_x,r_y,r_z\right]$, can be succinctly characterized as the linear combination of Fourier-harmonic-based steering vectors as follows
\begin{equation}\label{eq:point_to_point_channel}
{
\begin{aligned}
h(\mathbf{s},\mathbf{r})=&\sum_{(l_x,l_y)\in\xi_R}\sum_{(m_x,m_y)\in\xi_S}
\\
&H_a(l_x,l_y,m_x,m_y)a_R(l_x,l_y,\mathbf{r})a_S^H(m_x,m_y,\mathbf{s}),
\end{aligned}
}
\end{equation}
where $(l_x, l_y) \in \xi_R$ and $(m_x, m_y) \in \xi_S$ represent the wavenumber indices of the receiver and transmitter along the $ x$- and $y$-axis, respectively. $\xi_R$ and $\xi_S$ represent the wavenumber domain that collects all the wavenumber indices at the receiver and the transmitter.  $H_a(l_x,l_y,m_x,m_y)$ is a collection of independent complex-Gaussian random variables, i.e., $\mathcal{CN} \left(0, \sigma^2(l_x,l_y,m_x,m_y)\right)$, also known as the random Fourier coefficients corresponding to the wavenumber domain.
Still referring to (\ref{eq:point_to_point_channel}), $a_R\left(l_x, l_y, {\bf r} \right)$ and $a_S\left(m_x, m_y, {\bf s} \right)$ represent the scalar Fourier harmonics, respectively, which are explicitly given by
\begin{subequations}
\begin{align}
a_R(l_x,l_y, {\bf r})&=e^{j(\frac{2\pi l_x}{L_{R,x}} r_x+\frac{2\pi l_y}{L_{R,y}} r_y+k_z (\frac{2\pi l_x}{L_{R,x}},\frac{2\pi l_y}{L_{R,y}})r_z)},
\\
a_S(m_x,m_y, {\bf s})&=e^{j(\frac{2\pi m_x}{L_{S,x}} s_x+\frac{2\pi m_y}{L_{S,y}} s_y+k_z (\frac{2\pi m_x}{L_{S,x}},\frac{2\pi m_y}{L_{S,y}})s_z)},
\end{align}
\end{subequations}
where $k_z(k_x,k_y)$ denotes the wavenumber propagating along the $z$-axis, which is given by
\begin{equation}
k_z(k_x,k_y)=\sqrt{k^2-k_x^2-k_y^2}=\sqrt{(\frac{2\pi f_c}{c})^2-k_x^2-k_y^2},
\end{equation}
where $k_{x}, k_y$ denote the wavenumber along the $x$- and $y$-axis, respectively. When $k_z$ is imaginary, the propagating wave $e^{j(k_xx+k_yy+k_zz)}$ attenuates exponentially along the $z$-axis, which is referred to as the evanescent wave. Excluding evanescent wave from our analysis, we obtain the wavenumber constraint $k_x^2+k_y^2\leq k^2$. 
With $k_i = \frac{2\pi l_i}{L_{R,i}}, i\in \left\{x,y\right\}$ for the receiver and $k_i = \frac{2\pi m_i}{L_{S,i}}, i\in \left\{x,y\right\}$ for the transmitter, the wavenumber domain $\xi_R$ and $\xi_S$ in~(\ref{eq:point_to_point_channel}) are restricted within 
\begin{subequations}\label{xi}
\begin{align}
\xi_R&=\{(l_x,l_y)\in {\mathbb{Z}}^2:(\frac{l_x\lambda}{L_{R,x}})^2+(\frac{l_y\lambda}{L_{R,y}})^2\leq1\},
\\
\xi_S&=\{(m_x,m_y)\in {\mathbb{Z}}^2:(\frac{m_x\lambda}{L_{S,x}})^2+(\frac{m_y\lambda}{L_{S,y}})^2\leq1\}.
\end{align}
\end{subequations}
Then, let us examine the channel response spanning from the $n_S$-th antenna element of the transmitter to the $n_R$-th antenna element of the receiver, denoted by $\left[{\bf H}\right]_{n_R, n_S} $, with ${\bf H}\in \mathbb{C}^{N_R\times N_S}$ representing the channel matrix.
By substituting $\mathbf{s}={\bf p}_{S,n_S} = \left[\delta n_{S,x}, \delta n_{S,x}, 0\right]$ and $\mathbf{r}={\bf p}_{R,n_R} = \left[x_0+n_{R,x}\delta, y_0+n_{R,y}\delta, z_0\right]$ into~(\ref{eq:point_to_point_channel}), the $\left(n_R, n_S\right)$-th entry of the HMIMO channel, i.e., $\left[{\bf H}\right]_{n_R, n_S}$, can be structured as (\ref{eq:H_n}), shown at the bottom of next page.
\begin{figure*}[b]
	\hrule
	\small
\begin{align}\label{eq:H_n}
\sum\limits_{\begin{subarray}{l} 
	\left( {{l_x},{l_y}} \right) \in {\xi _R} \\ 
	\left( {{m_x},{m_y}} \right) \in {\xi _S} 
\end{subarray}}  {{H_a}\Big( {{l_x},{l_y},{m_x},{m_y}} \Big)\exp \{ j\Big[\frac{{2\pi {l_x}({n_{R,x}}\delta + x_0)}}{{{L_{R,x}}}} +\frac{{2\pi {l_y}({n_{R,y}}\delta + y_0)}}{{{L_{R,y}}}} + {k_z}\left( {\frac{{2\pi {l_x}}}{{{L_{R,x}}}},\frac{{2\pi {l_y}}}{{{L_{R,y}}}}} \right){z_0} - \frac{{2\pi {m_x}{n_{S,x}}\delta }}{{{L_{S,x}}}} - \frac{{2\pi {m_y}{n_{S,y}}\delta }}{{{L_{S,y}}}}\Big]\} } .
\end{align}
\end{figure*}
Given that $H_a(l_x,l_y,m_x,m_y)\sim\mathcal{CN}(0,\sigma^2(l_x,l_y,m_x,m_y))$ and $x_0,y_0,z_0\in\mathbbm{R}$, we obtain the statistical equivalence 
\begin{align}
&H_a(l_x,l_y,m_x,m_y)\sim \ H_a(l_x,l_y,m_x,m_y)\times
\nonumber \\
& \quad \exp\{j\Big[\frac{2\pi l_x x_0}{L_{R,x}}+\frac{2\pi l_y y_0}{L_{R,y}}+k_z \Big(\frac{2\pi l_x}{L_{R,x}},\frac{2\pi l_y}{L_{R,y}}\Big)z_0\Big]\}.
\end{align}
Therefore, the terms containing $x_0$, $y_0$ and $z_0$ in ({\ref{eq:H_n}}) can be removed, and $\left[{\bf H}\right]_{n_R, n_S}$ can be further recast to
\begin{align}\label{eq:H_n_scalar}
&\left[{\bf H}\right]_{n_R, n_S} = 
\sum_{\left(l_x, l_y\right) \in \xi_R}
\sum_{\left(m_x, m_y\right) \in \xi_S}
H_a\left(l_x,l_y,m_x,m_y\right) \times  \nonumber\\
&\resizebox{0.9\hsize}{!}{$
\exp \Bigg\{j\Big(\frac{2\pi l_xn_{R,x}\delta}{L_{R,x}}+\frac{2\pi l_yn_{R,y}\delta}{L_{R,y}}-\frac{2\pi m_xn_{S,x}\delta}{L_{S,x}}-\frac{2\pi m_yn_{S,y}\delta}{L_{S,y}}\Big)\Bigg\}. $}
\end{align}
We denote $\mathbf{x}_p\in \mathbb{C}^{N_S\times1}$ by the pilot signal from the transmitter in the $p$-th time slot. The corresponding received pilot signal in the $p$-th time slot ${\bf y}_p \in \mathbb{C}^{{N_{RF}}\times 1}$ is given by
\begin{equation}
\begin{aligned}
\mathbf{y}_p=\mathbf{CHx}_p+\mathbf{n},
\end{aligned}
\end{equation}
where $\mathbf{C}\in\mathbbm{C}^{N_{RF}\times N_R}$ denotes the combining matrix which reduces the dimension of the received signal from $N_R$ to $N_{RF}$, and $\mathbf{n}\sim \mathcal{CN} \left({\bf 0}, \sigma_n^2 {\bf I}_{N_{RF}}\right)$ denotes the additive white Gaussian noise (AWGN). Let $P$ denote the pilot length for channel estimation. Thus, the overall transmitted pilot can be aggregated into $\mathbf{X}=[\mathbf{x}_1,\mathbf{x}_2,...,\mathbf{x}_{P}]$, and the overall received pilot can be given by $\mathbf{Y}=[\mathbf{y}_1,\mathbf{y}_2,...,\mathbf{y}_{P}]$. { When the antenna spacing $\delta$ is less than half-wavelength, the mutual coupling effect (i.e. the voltage on each antenna element will
induce current on its nearby antenna elements) can not be ignored. In this case, the channel matrix $\mathbf{H}$ is recast to $\mathbf{M}_R\mathbf{HM}_S$, where $\mathbf{M}_R$ and $\mathbf{M}_S$ denote the mutual coupling matrix at the receiver and transmitter, respectively. These matrices can be determined numerically once the antenna array is manufactured~\cite{mutualcoupling}. Given that $\mathbf{M}_R$ can be included in $\mathbf{C}'\equiv\mathbf{CM}_R$, and $\mathbf{M}_S$ can be included in $\mathbf{x}_p'\equiv\mathbf{M}_S\mathbf{x}_p$, the mutual coupling effect does not affect the sparsity structure of the channel matrix. }

\subsection{{Sparse Representation of the HMIMO Channel}} \label{subsec2b}
The scalar representation in~(\ref{eq:H_n_scalar}) can be further refined into its matrix counterpart
\begin{equation}\label{wavenumberdomain}
\begin{aligned}
\mathbf{H}=\mathbf{\Psi}_R\mathbf{H}_a{\mathbf{\Psi}}^H_S,
\end{aligned}
\end{equation}
    where $\mathbf{\Psi}_R\in\mathbb{C}^{N_R\times \vert\xi_R\vert}$ and ${\mathbf{\Psi}}_S\in\mathbb{C}^{N_S\times \vert\xi_S\vert}$ are the wavenumber-domain  sparsifying basis at the receiver and transmitter, respectively. Upon denoting  $l\triangleq (l_x,l_y)\in\xi_R$, $m\triangleq (m_x,m_y)\in\xi_S$, the $\left( n_R,l\right) $-th entry of $[\mathbf{\Psi}_R]$ and the $\left( {n_S,m}\right) $- entry of $[\mathbf{\Psi}_S]$ are give by 
\begin{subequations}\label{Psi}
\begin{align}
[\mathbf{\Psi}_R]_{n_R,l}&=\frac{1}{\sqrt{N_R}}\exp\Big\{j\left(\frac{2\pi l_xn_{R,x}\delta}{L_{R,x}}+\frac{2\pi l_yn_{R,y}\delta}{L_{R,y}}\right)\Big\}, \\
[\mathbf{\Psi}_S]_{n_S,m}&=\frac{1}{\sqrt{N_S}}\exp\Big\{j\left(\frac{2\pi m_xn_{S,x}\delta}{L_{S,x}}+\frac{2\pi m_yn_{S,y}\delta}{L_{S,y}}\right)\Big\}.
\end{align}
\end{subequations}
Still referring to (\ref{wavenumberdomain}), $\mathbf{H}_a\in\mathbb{C}^{\vert\xi_R\vert\times\vert\xi_S\vert}$ is known as the  wavenumber-domain channel matrix \cite{fourierantennaefficiency}, the dimension of which is irrespective to the antenna spacing with fixed array sizes \cite[Eq. (19)]{fourier}. The physical meaning of $\left[\mathbf{H}_a\right]_{l,m}$ can be interpreted as the complex gain along the propagation direction of the $m$-th transmitted wavenumber and the propagation direction of the $l$-th received wavenumber. Each coefficient $\left[\mathbf{H}_a\right]_{l,m}$ follows independent complex-Gaussian distribution $\mathcal{CN}(0,\sigma^2(l_x,l_y,m_x,m_y))$. 
By applying the separable model presented in~\cite{fourier}, the variance $\sigma^2(l_x,l_y,m_x,m_y)$ can be further decoupled into $\sigma^2(l_x,l_y,m_x,m_y)=\sigma_R^2(l_x,l_y)\sigma_S^2(m_x,m_y)$,
where the variance factors $\sigma_R^2\left(l_x,l_y\right)$ and $\sigma_S^2\left(m_x,m_y\right)$ are determined by the integration on the spectral factors $A^2_S(\theta_S,\phi_S)$  and $A^2_R(\theta_R,\phi_R)$ in the corresponding wavenumber domain, as detailed in \cite[Eq. (30)]{fourier}. The spectral factor observed at the transmitter side $A^2_S(\theta_S,\phi_S)$ characterizes the average power transmitted in every direction, and the spectral factor observed as the receiver side $A^2_R(\theta_R,\phi_R)$ characterizes the average power  received in every direction. Let $\bm{\sigma}_R\in\mathbb{R}^{\vert\xi_R\vert\times1}$ collect the variances $\sigma_R^2(l_x,l_y)$ and $\bm{\sigma}_S\in\mathbb{R}^{\vert\xi_S\vert\times1}$ collect the variances $\sigma_S^2(m_x,m_y)$, and $\mathbf{H}_a$ can be further structured as
\begin{equation}\label{ha}
\mathbf{H}_a=\rm{diag}(\bm{\sigma}_R)\mathbf{W}\rm{diag}(\bm{\sigma}_S),
\end{equation}
where each entry of $\mathbf{W}\in\mathbbm{C}^{\vert\xi_R\vert\times\vert\xi_S\vert}$ follows a complex Gaussian distribution with zero mean and unit variance. Consequently, the channel matrix $\mathbf{H}$ has a decomposed form
\begin{equation}
\mathbf{H}=\mathbf{\Psi}_R\rm{diag}(\bm{\sigma}_R)\mathbf{W}\rm{diag}(\bm{\sigma}_S){\mathbf{\Psi}}^H_S.
\end{equation}

Considering realistic wireless communications, radio wave propagation is typically clustered around some significant propagation directions \cite{cluster}. In this case, the spectral factors $A^2_R(\theta_R,\phi_R)$ and $A^2_S(\theta_S,\phi_S)$ can be modeled by the mixture of VMF distribution \cite{spatialcharacter}. Specifically, supposing there are $N_c$ clusters in the propagation environment, the spectral factor $A^2_I(\theta_I,\phi_I)$ ($I\in\{R,S\}$) can be modeled as
\begin{equation}
\begin{aligned}
A^2_I(\theta_I,\phi_I)=\sum^{N_c}_{i=1}w_{I,i}p_{I,i}(\theta_I,\phi_I),
\end{aligned}
\end{equation}
where $\sum^{N_c}_{i=1}w_{I,i}=1$ are the normalization factors, and $p_{I,i}$ represents the probability density function of the three-dimensional von Mises-Fisher (VMF) distribution \cite{fourierantennaefficiency}
\begin{align}
{p_{I,i}}\left( {{\theta _I},{\phi _I}} \right) =& \frac{{{\alpha _{I,i}}}}{{4\pi {\text{sinh}}({\alpha _{{\text{I}},{\text{i}}}})}}\exp \left\{ {{\alpha _{I,i}}} \right.\left[ {\sin \;{\theta _R}\sin \;{\theta _{I,i}}  } \times \right. \nonumber\\
&  \left. {\left. {  \cos \;({\phi _I} - {\phi _{I,i}}) + \cos \;{\theta _I}\cos \;{\theta _{I,i}}} \right]} \right\}.
\end{align}
Here, $\{\phi_{S,i},\theta_{S,i}\}$ are the zenith angle of departure (ZoD) and the azimuth angle of departure (AoD) of the $i$-th cluster, respectively. $\{\phi_{R,i},\theta_{R,i}\}$ are the zenith angle of arrival (ZoA) and the azimuth angle of arrival (AoA) of the $i$-th cluster, respectively. $\alpha_{R,i}>0$ and $\alpha_{S,i}>0$ represent the concentration parameters of each cluster which are determined by the angular spread (AS). We denote $\delta_{ASD}(^\circ)$ by the AS observed at the transmitter side and $\delta_{ASA}(^\circ)$ by the AS observed at the receiver side. For $\delta_{ASA}<21^{\circ}$ and $\delta_{ASD}<21^{\circ}$, the relation between the AS and the concentration parameter is given by~\cite{fourierantennaefficiency}
\begin{equation}
\alpha_{R,i}=\frac{212.9^2}{\delta^2_{ASA}},\ \alpha_{S,i}=\frac{212.9^2}{\delta^2_{ASD}}.
\end{equation}
In realistic propagation environment, $\delta_{ASA}$ and $\delta_{ASD}$ can be influenced by carrier frequency, antenna gain, and half-power beam bandwidth \cite{AS}. When the HMIMO system operates at high frequency, e.g., millimeter-wave frequency, the antennas are usually highly directional with narrow beam. With the distance between the transmitter and the receiver being 200~m, $\delta_{ASA}$ and $\delta_{ASD}$ are usually not larger than $18^{\circ}$ \cite{AS}, implying $\alpha_{R,i}>140$ and $\alpha_{S,i}>140$. Inspired by the two-sigma rule for Gaussian distribution \cite{twosigma}, we define the number of significant entires in $\mathbf{H}_a$ as the minimum number of entries that collectively account for at least 95.44\% of the total energy. In this case, the significant entries of $\mathbf{H}_{a}$ are typically sparse. Glancing at (\ref{ha}), the sparsity level of $\mathbf{H}_a$ can be approximated by the product of the sparsity levels of $\bm{\sigma}_R$ and $\bm{\sigma}_S$. This implies that $\mathbf{H}_a$ may become sparse even in cases where $\bm{\sigma}_R$ and $\bm{\sigma}_S$ individually do not demonstrate sparsities. 

\begin{figure}[t]
\centering
\includegraphics[width=0.5\textwidth]{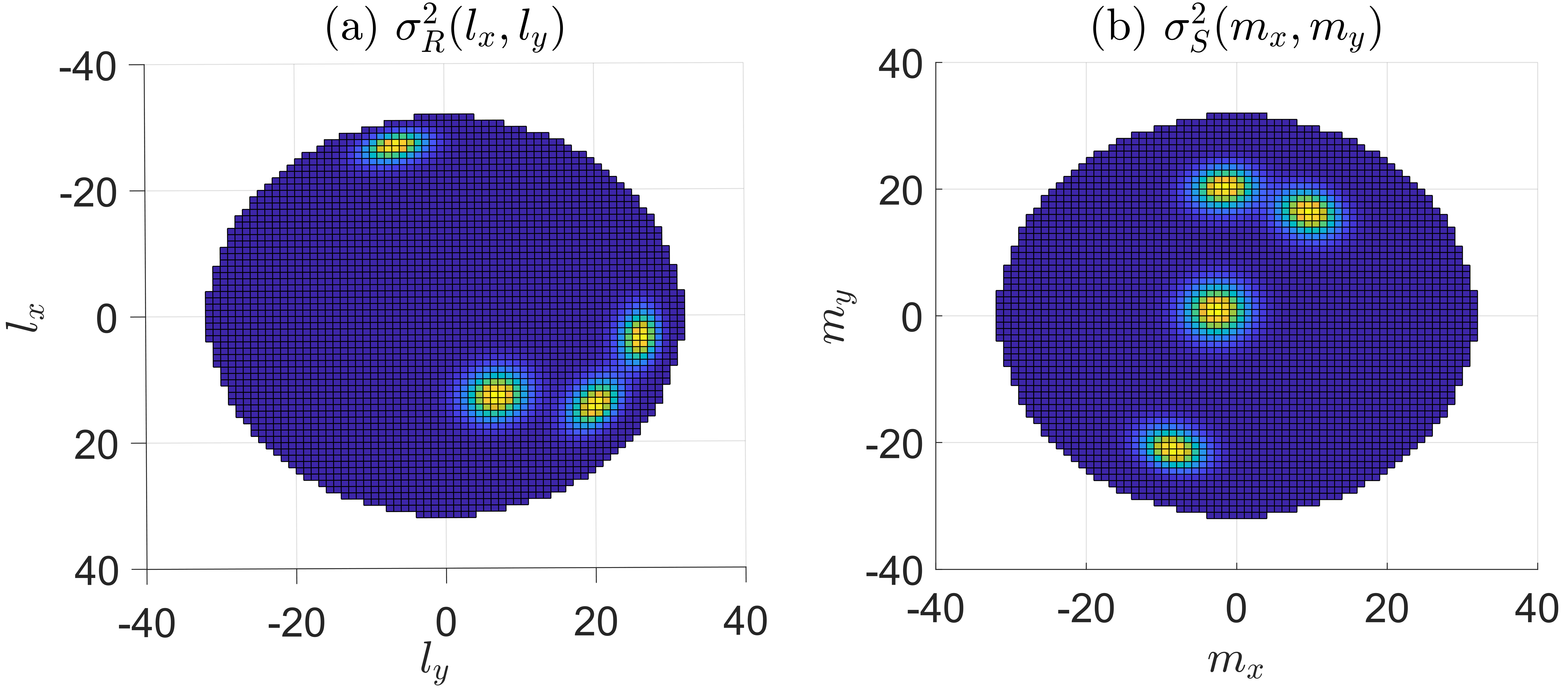}
\caption{Variances ${\sigma}^2_R(l_x,l_y)$ and $\sigma^2_S(m_x,m_y)$ structured in the wavenumber domain. The parameters are $L_{R,x}=L_{R,y}=L_{S,x}=L_{S,y}=32.25\lambda$, $N_c=4$, and $\alpha_{R,i}=\alpha_{S,i}=140$.  $\theta_{R,i}$ and $\theta_{S,i}$ are randomly sampled within $[0^\circ,90^{\circ}]$. $\phi_{R,i}$ and $\phi_{S,i}$ are randomly sampled within $[0^\circ,360^{\circ}]$.}
	\label{fig1}
 \vspace{-1em}
\end{figure}

Fig.~\ref{fig1} demonstrates variances $\sigma^2_R(l_x,l_y)$ and $\sigma^2_S(m_x,m_y)$ structured in the wavenumber domain. Variances $\sigma^2_R(l_x,l_y)$ and $\sigma^2_S(m_x,m_y)$ visualize the angular power spectrum of the HMIMO channel observed at the receiver side and the transmitter side, respectively. In particular, each grid in Fig.~\ref{fig1}(a) and Fig.~\ref{fig1}(b) represents the average power received or transmitted
along its corresponding propagation direction, i.e., $k_{R,x}=2\pi l_x/L_{R,x}$,  $k_{R,y}=2\pi l_y/L_{R,y}$, $k_{S,x}=2\pi m_x/L_{S,x}$, and $k_{S,y}=2\pi m_y/L_{S,y}$.

\section{Sparse HMIMO Chanel Estimation} 

The goal of channel estimation is to estimate HMIMO channel $\mathbf{H}$ from the received pilot $\mathbf{Y}\in\mathbbm{C}^{N_{RF}\times P}$ and the transmitted pilot $\mathbf{X}\in\mathbbm{C}^{N_{S}\times P}$. By harnessing the sparse representation in  (\ref{wavenumberdomain}), the channel estimation problem can be formulated as 
\begin{equation}\label{problem}
\hat{\mathbf{H}}_a=\arg\min_{\hat{\mathbf{H}}_a}\{\ \vert\vert\mathbf{C\Psi}_R{\mathbf{\hat{H}}}_a\mathbf{\Psi}^H_S\mathbf{X}-\mathbf{Y}\vert\vert_F^2\ \},
\end{equation}
where $\mathbf{\hat{H}}_a\in\mathbbm{C}^{\vert\xi_R\vert\times \vert\xi_S\vert}$ is the sparse matrix to be estimated. 
Then, the OMP algorithm is extended to function efficiently in the wavenumber domain  for accurately recovering sparse HMIMO channel $\mathbf{\hat{H}}_a$.


For ease of exposition, we define $\mathbf{A}=\mathbf{C\Psi}_R$ and $\mathbf{B}=\mathbf{\Psi}^H_S\mathbf{X}$. The $i$-th column of $\mathbf{A}$ is denoted by $\mathbf{a}_i\in\mathbbm{C}^{N_{RF}\times1}$ ($1\leq i\leq \vert\xi_R\vert$), and the $j$-th row of $\mathbf{B}$ is denoted by $\mathbf{b}^T_j\in\mathbbm{C}^{1\times P}$ ($1\leq j\leq \vert\xi_S\vert$). Firstly, in the $u$-th iteration, 
we search the $i^{\left( u \right)}$-th column of $\mathbf{A}$ and the $j^{\left( u \right)}$-th row of $\mathbf{B}$ such that the matrix $\mathbf{a}_{i^{(u)}}\mathbf{b}^T_{j^{(u)}}$ is most relevant to the residual matrix $\mathbf{Y}_{res}$, after which the entry indexed by $(i^{(u)},j^{(u)})$ is added into the sparse support set $\mathcal{H}$. Then, we minimize the residual with the existing sparse support by solving the problem
\begin{subequations}\label{lsproblem}
\begin{align}
&\arg\min\limits_{\mathbf{w}}\ \vert\vert\ \mathbf{Y}-\sum_{u'=\rm{1}}^{u}w_{u'}\ \mathbf{a}_{i^{(u')}}\mathbf{b}^T_{j^{(u')}}\ \vert\vert^2_F
\\
\label{lsproblem2}\equiv&\arg\min\limits_{\mathbf{w}}\  \rm{tr}\mathit{(\mathbf{Y}\mathbf{Y}^H)-\mathbf{f}^T\mathbf{w}-\mathbf{f}^H\mathbf{w}^*+\mathbf{w}^T\mathbf{F}\mathbf{w}^*},
\end{align}
\end{subequations} 
where $\mathbf{w}=[w_1,w_2,...,w_u]^T$ denotes the solution to problem (\ref{lsproblem}), with $\mathbf{f}\in\mathbbm{C}^{u\times 1}$ and $\mathbf{F}\in\mathbbm{C}^{u\times u}$ given by
\begin{equation}
[\mathbf{f}]_n=\rm{tr}\mathit{(\mathbf{a}_{i^{(n)}}\mathbf{b}^T_{j^{(n)}}\mathbf{Y}^H)},\ \rm{1}\mathit{\leq n\leq u},
\end{equation}
\begin{equation}
[\mathbf{F}]_{\mathit{m,n}}=\rm{tr}(\mathit{\mathbf{a}_{i^{(m)}}\mathbf{b}^T_{j^{(m)}}\mathbf{b}^*_{j^{(n)}}\mathbf{a}^H_{i^{(n)}}}),\ \rm{1}\mathit{\leq m,n\leq u}.
\end{equation}
Taking the gradient of (\ref{lsproblem2}), the solution to problem (\ref{lsproblem}) is
\begin{equation}\label{solution}
[w_1,w_2...w_u]^T=(\mathbf{F}^{-1}\mathbf{f})^*.
\end{equation}
At the end of each iteration, the residual matrix $\mathbf{Y}_{res}$ is updated with the existing sparse support $\mathcal{H}$ and (\ref{solution}). Based on the results of all iterations, we obtain the wavenumber-domain matrix  $\mathbf{\hat{H}}_a$, and thus the HMIMO channel $\mathbf{\hat{H}}$ can be reconstructed according to the pattern presented in (\ref{wavenumberdomain}). 
The overall process for recovering $\mathbf{\hat{H}}$ is summarized in Algorithm~\ref{alg}.

\begin{algorithm}[t]
{\small
\caption{Wavenumber-Domain OMP (WD-OMP)}
{\bf Input:} Transmitted pilot $\mathbf{X}$; received pilot $\mathbf{Y}$; combining matrix $\mathbf{C}$, wavenumber-domain sparsifying matrix $\mathbf{\Psi}_R$ and $\mathbf{\Psi}_S$, number of iterations $U_{iter}$.

{\bf Output:} HMIMO channel matrix $\mathbf{\hat{H}}$.
\begin{algorithmic}[1]\label{alg}
\STATE Initialize $\mathbf{A}=\mathbf{C\Psi}_R$, $\mathbf{B}=\mathbf{\Psi}^H_S\mathbf{X}$, $\mathbf{\hat{H}}_a=\mathbf{O}$, $\mathbf{Y}_{res}=\mathbf{Y}$, $\mathcal{H}=$$\varnothing$.
\FOR{$u\in \{1,2,\dots,U_{iter}\}$}
\STATE $(i^{(u)},j^{(u)})\longleftarrow\rm{arg}$ $\max\limits_{(i,j)\notin\mathcal{H}}\frac{\vert\langle\mathbf{a}_{i}\mathbf{b}^T_{j},\mathbf{Y}_{res}\rangle\vert}{\vert\vert\mathbf{a}_{i}\mathbf{b}^T_{j}\vert\vert_F \vert\vert\mathbf{Y}_{res}\vert\vert_F}$.
\STATE Update the sparse support: $\mathcal{H}\longleftarrow \mathcal{H}$ $\cup$ $(i^{(u)},j^{(u)})$.
\STATE Solve (\ref{lsproblem}) with (\ref{solution}) to minimize the residual.
\STATE Update the residual: $\mathbf{Y}_{res}\longleftarrow\mathbf{Y}-\sum_{u'=1}^{u}w_{u'}\ \mathbf{a}_{i^{(u')}}\mathbf{b}^T_{j^{(u')}}$.
\ENDFOR
\FOR{$u \in\{ 1,2,\dots,U_{iter}\}$}
\STATE $\mathbf{\hat{H}}_a(i^{(u)},j^{(u)})=w_{u}$.
\ENDFOR
\STATE $\mathbf{\hat{H}}=\mathbf{\Psi}_R\mathbf{\hat{H}}_a\mathbf{\Psi}_S$.
 \RETURN $\mathbf{\hat{H}}$
\end{algorithmic}}
\end{algorithm}

\begin{figure*}[t]
	\centering
	\begin{subfigure}[b]{0.32\linewidth}
		\centering
		\includegraphics[width=0.98\linewidth]{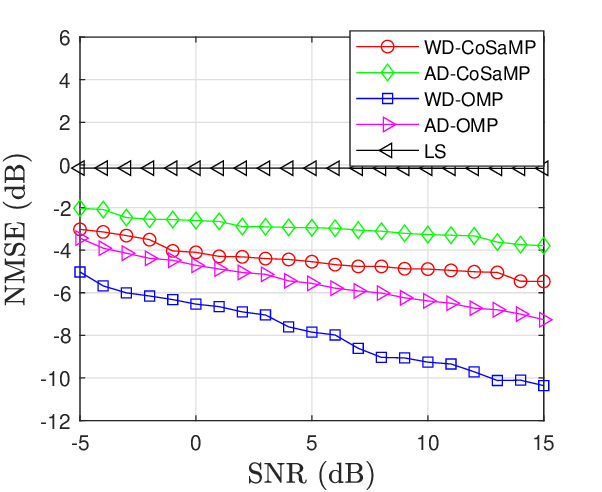}
		\caption{}
		\label{fig:sub1}
	\end{subfigure}
	\hfill
	\begin{subfigure}[b]{0.32\linewidth}
		\centering
		\includegraphics[width=0.98\linewidth]{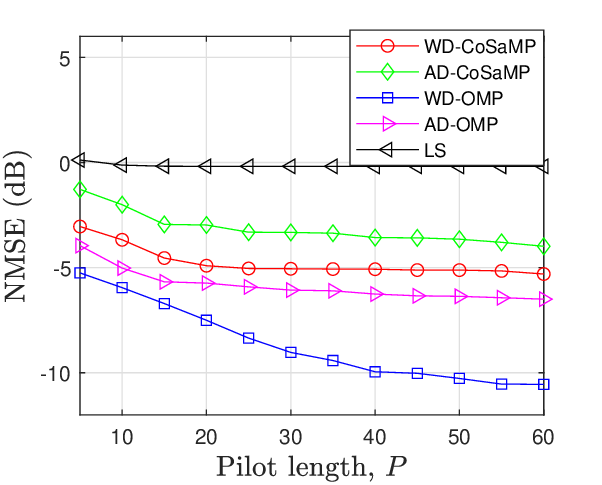}
		\caption{}
		\label{fig:sub2}
	\end{subfigure}
	\hfill
	\begin{subfigure}[b]{0.32\linewidth}
		\centering
		\includegraphics[width=0.98\linewidth]{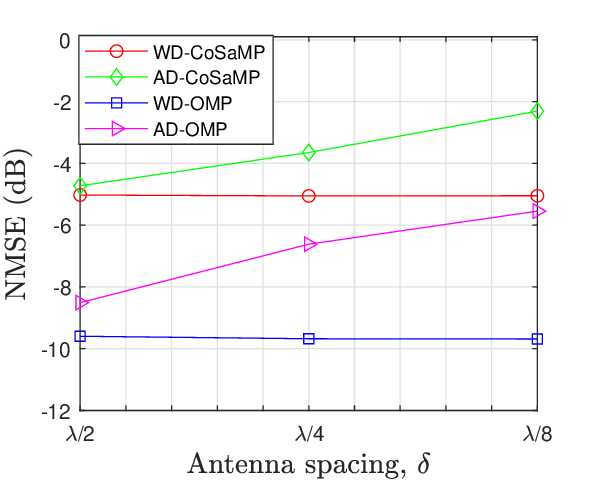}
		\caption{}
		\label{fig:sub3}
	\end{subfigure}
	\caption{(a) NMSE performance versus SNR. (b) NMSE performance versus the pilot length $P$. (c) NMSE performance versus antenna spacing $\delta$.
 }\label{fig:combined}
\end{figure*}

\section{Simulation Results}
In this section, simulation results are provided to evaluate the channel estimation performance of the proposed WD-OMP algorithm for the HMIMO system of interest. The transmitter has $N_S = 5 \times 5$ and the receiver has $ N_R = 65 \times 65$ antenna elements. The antenna  spacing is $\delta = \lambda /4$. The receiver includes $N_{RF} = 64$ RF chains while the transmitter only has one RF chain for processing baseband signals. The considered HMIMO system operates at 30~GHz. It is assumed that the number of clusters is $N_c = 2$. Each channel realization involves sampling $\theta_{R,i}$ and $\theta_{S,i}$ from a uniform distribution $\mathcal{U}(0,\pi / 2)$, and sampling $\phi_{R,i}$ and $\phi_{S,i}$ from a uniform distribution $\mathcal{U}(0,2\pi)$. The concentration parameters of both transmitter and receiver side are identical, i.e., $\alpha_{S,i} = \alpha_{R,i} =140$ (see Sec. \ref{subsec2b}). Furthermore, we concentrate on the normalized mean square
error (NMSE) performance for quantifying the proposed algorithm,  which is defined as ${\text{NMSE}} = \mathbb{E}\left\{ {\left\| {{\mathbf{\hat H}} - {\mathbf{H}}} \right\|_2^2} \right\}/\mathbb{E}\left\{ {\left\| {\mathbf{H}} \right\|_2^2} \right\}$. The signal-to-noise rate (SNR) is SNR $=\frac{\vert\vert\mathbf{CHX}\vert\vert^2_2}{PN_{RF}\sigma^2_n}$. Each entry of the pilot sequence $\mathbf{X}$ is independently selected from $\frac{1}{\sqrt{N_S}}\{1,-1\}$ with equal probability, and each entry of the combining matrix $\mathbf{C}$ follows i.i.d. $\mathcal{CN}(0,1/N_R)$ \cite{vtguo}. { The Rayleigh distance, calculated by $2D^2/\lambda$ where $D$ represents the maximum aperture, is 10.24~m for the received array and 0.04~m for the transmitted array. Therefore, the near-field effect is negligible and can be ignored in the scenario at hand.}

The following benchmark schemes are employed for comparison. i) Least square (LS) is applied for HMIMO channel estimation. ii) Angular-domain OMP (AD-OMP): The HMIMO channel $\mathbf{H}$ is estimated using OMP \cite[Algorithm~1]{CS} based on the angular-domain representation in \cite{AD}. iii) Angular-Domain CoSaMP (AD-CoSaMP): The HMIMO channel $\mathbf{H}$ is estimated using CoSaMP Algorithm \cite[Algorithm~2]{CS} based on the angular-domain representation. iv) Wavenumber-Domain CoSaMP (WD-CoSaMP): The HMIMO channel $\mathbf{H}$ is estimated using CoSaMP Algorithm based on the wavenumber-domain representation in (\ref{wavenumberdomain}).




Fig.~\ref{fig:combined}(a) illustrates the NMSE performance versus varying SNR. The pilot length $P$ is 32 and the compression ratio $\frac{PN_{RF}}{N_SN_R}$ is maintained at 0.02. As can be observed, for a low compression ratio, LS is rendered inefficient due to a high degree of correlation among the entries of the measurement matrix. Similarly, the performance of CoSaMP is limited under conditions of low compression ratio.
This limitation arises from the algorithm's iterative process, which involves incorporating a group of entries with the highest correlation into the support set, leading to ambiguity in detection.

Fig.~\ref{fig:combined}(b) plots the NMSE performance versus the pilot length $P$ when SNR = 10~dB.
The pilot length $P$, extending from 5 to 60, leads to an increase in the compression ratio $\frac{PN_{RF}}{N_SN_R}$, from 0.003 to 0.036. An evident performance saturation is observed with an increment of $P$. This can be attributed to the fact that both the channel matrices filtered by the angular-domain and the wavenumber-domain sparsifying bases exhibit relatively weak sparsity, rendering the CS-based algorithms less effective, irrespective of the pilot length. 

Next, we intend to examine how antenna spacing influences channel estimation performance attained by angular-domain and wavenumber-domain approaches.
Illustrated in Fig.~\ref{fig:combined}(c) is the NMSE performance resultant from different antenna spacing configurations, set against a context of a fixed UPA size, i.e., $L_{R,x}=L_{R,y}=16\lambda$ at the receiver and $L_{S,x}=L_{S,y}=\lambda$ at the transmitter. As transpired in Fig.~\ref{fig:combined}(c), the wavenumber-domain approach outperforms its angular-domain counterpart in terms of NMSE performance. When the antenna spacing is decreased to being below half a wavelength, a discernible deterioration in the angular-domain approach is spotted, while the NMSE performance achieved by wavenumber-domain  remains stable. {
{ This can be understood owing to the following facts. On the one hand, the size of the AD channel matrix is $N_R N_S= {L_{S,x}L_{S,y}L_{R,x}L_{R,y}}/{\delta^4}$~\cite{AD}, which increases significantly as the antenna spacing decreases when the array aperture is fixed. Nevertheless, reducing the antenna spacing and introducing additional grids in the AD channel matrix does not contribute to improving spatial resolution. This is because the spatial resolution of the array is solely determined by the size of the array aperture~\cite{fourier}. These additional grids, however, increase the possibility of detection ambiguity. On the other hand, the size of the WD channel matrix is $\vert \xi_R \vert  \vert \xi_S \vert$, which remains constant despite variations in antenna spacing when the array aperture remains constant. Consequently, as the antenna spacing decreases, AD-based approaches are more prone to resulting in detection ambiguity, whereas WD-based approaches still maintain their effectiveness.}


\section{Conclusion}
In this work, we investigate the sparse HMIMO channel estimation in the wavenumber domain, based on a customized wavenumber-domain sparsifying basis. Then, the WD-OMP algorithm is proposed, specifically tailored for efficiently solving the formulated CS problem.
Finally, simulation results demonstrate the proposed algorithm's capability of delivering favorable accuracy across a range of practical compression ratios and SNR levels. Notably, the proposed wavenumber-domain approach exhibits great robustness in the presence of  variations in antenna spacing for a fixed array aperture.


\vspace{3em}

\bibliographystyle{IEEEtran}
\bibliography{referencebox}

\end{document}